\documentclass[journal=jacsat,manuscript=article]{achemso}
\usepackage[utf8]{inputenc}
\usepackage{amsmath}
\usepackage{amsmath,amssymb}
\usepackage{graphicx}
\usepackage{caption}
\usepackage{color}
\usepackage{dcolumn}
\usepackage{bm}
\usepackage{float}
\usepackage{subfig}
\usepackage{url}

\author{Markus Meuwly}\email{m.meuwly@unibas.ch}
\affiliation{Department of Chemistry, University of Basel,
  Klingelbergstrasse 80, CH-4056 Basel, Switzerland}
\affiliation{Department of Chemistry, Brown University, Providence RI,
  USA}

\author{Martin Karplus}\email{marci@tammy.harvard.edu}
\affiliation{Department of Chemistry, Harvard University, USA}
\affiliation{Laboratoire de Chimie Biophysique, ISIS, Universit\'{e}
  de Strasbourg, 67000 Strasbourg, France}

\title{The Functional Role of the Hemoglobin-Water Interface}

\begin{document}

\date{\today}

\begin{abstract}
The interface between hemoglobin (Hb) and its environment, in
particular water, is of great physiological relevance. Here, results
from {\it in vitro}, {\it in vivo}, and computational experiments
(molecular dynamics simulations) are summarized and put into
perspective. One of the main findings from the computations is that
the stability of the deoxy, ligand-free T-state (T$_0$) can be
stabilized relative to the deoxy R-state (R$_0$) only in sufficiently
large simulation boxes for the hydrophobic effect to manifest
itself. This effect directly influences protein stability and is
operative also under physiological conditions. Furthermore, molecular
simulations provide a dynamical interpretation of the Perutz model for
Hb function. Results from experiments using higher protein
concentrations and realistic cellular environments are also
discussed. One of the next great challenges for computational studies,
which as we show is likely to be taken up in the near future, is to
provide molecular-level understanding of the dynamics of proteins in
such crowded environments.
\end{abstract}

\section{Introduction}
The human red blood cell (RBC, erythrocyte) contains a complex aqueous
solution of hemoglobin, nonhemoglobin proteins, lipids, glucose,
electrolytes (mainly K$^+$, Na$^+$, Cl$^-$, HCO$_3^-$, and phosphates)
and other components.  Approximately 97 \% of its volume is occupied
by water and hemoglobin and that of intracellular water alone of 72
\%.\cite{levin:1976} Hence, the main interface between hemoglobin and
its intracellular environment is with water.  Thus, it is essential to
be able to describe the interaction between Hb and water for
understanding the physiological function of Hb in RBCs.\\

\noindent
In this contribution an overview of the current knowledge of the
Hb/water interface is provided with an emphasis on the structure and
dynamics of the protein and its environment. The solution properties
of hemoglobin were already reviewed by Antonini and Brunori in their
book published fifty years ago.\cite{antonini:1971} At the time the
solution structure of the protein, its oligomerization state and the
behaviour in solutions with different ionic strengths were of
particular interest. A protein's solvent environment and its chemical
composition can change its effects from acting as plasticizers
(e.g. water) to being stabilizers (e.g. trehalose or glycerol). At a
molecular level this is further complicated by the fact that protein
side chains can switch between conformational substates and that
certain such motions can be prevented. For example, for lysozyme it
was shown experimentally that the internal dynamics is activated when
the environment changes from pure glycerol (which is a stabilizer) by
increasing the level of hydration, because water is a
plasticizer.\cite{paciaroni:2002} More generally it has been found
that protein dynamics is {\it slaved} to solvent
fluctuations.\cite{fenimore:2002} Increased hydration has also been
found to affect internal motions.\cite{smith:1989} Hence, it is
essential to better understand the interplay between solvent structure
and dynamics and the protein dynamics coupled to it and whether this
influences the biological function in a physiological context.\\

\noindent
A first dynamical transition of the internal dynamics of proteins
between ``low amplitude motion'' and more ``diffusive motion'' occurs
between 180 K and 220 K\cite{doster:1989} although for the protein
thaumatin a transition at 110 K has been reported.\cite{gruner:2011}
This ``glass transition'' was investigated by using quasi-inelastic
neutron scattering experiments or M\"ossbauer
spectroscopy.\cite{parak:2002} Experiments on Hb in RBCs have reported
an ``elastomeric transition'' between a gel- and a fluid-like
phase.\cite{artmann:1998} At a temperature of 310 K, human RBCs were
found to undergo a sudden change from blocking to passing through
micropipettes. This behavior was associated with a gel-to-fluid phase
transition\cite{artmann:1998} which is also related to the known
increase in viscosity of highly concentrated Hb. Subsequent
experiments on RBCs from different organisms revealed that such a
transition involves Hb in all cases and that, more surprisingly, the
transition temperature is correlated with the body temperature of the
respective species.\cite{stadler:2008}\\

\noindent
It was hypothesized that the drop in viscosity with subsequent changes
for cellular passage through micropipettes is caused by protein
aggregation.\cite{artmann:1998} Concomitantly, Hb shows a pronounced
loss of its $\alpha-$helical content at body temperature, not only for
human Hb, but also for several other species.\cite{zerlin:2007} It was
proposed\cite{digel:2006} that this is due to an increased amplitude
of the sidechain motions. This triggers unfolding of the helical
structure accompanied by an increase in surface hydrophobicity, which
eventually leads to protein aggregation. Experimentally, this was
investigated by temperature-dependent incoherent quasielastic neutron
scattering on whole red blood cells.\cite{stadler:2008} Such
experiments are able to separate global and internal protein
motions. It was found that the amplitudes of protein side-chain
motions increased close to the body temperature of different species,
such as monotremes (305 K) and humans (311 K) with the difference due
to amino acid substitutions.\\

\section{The Role of Water for the Stability of Hemoglobin}
\begin{figure}[H]
\centering \includegraphics[width=0.9\textwidth]{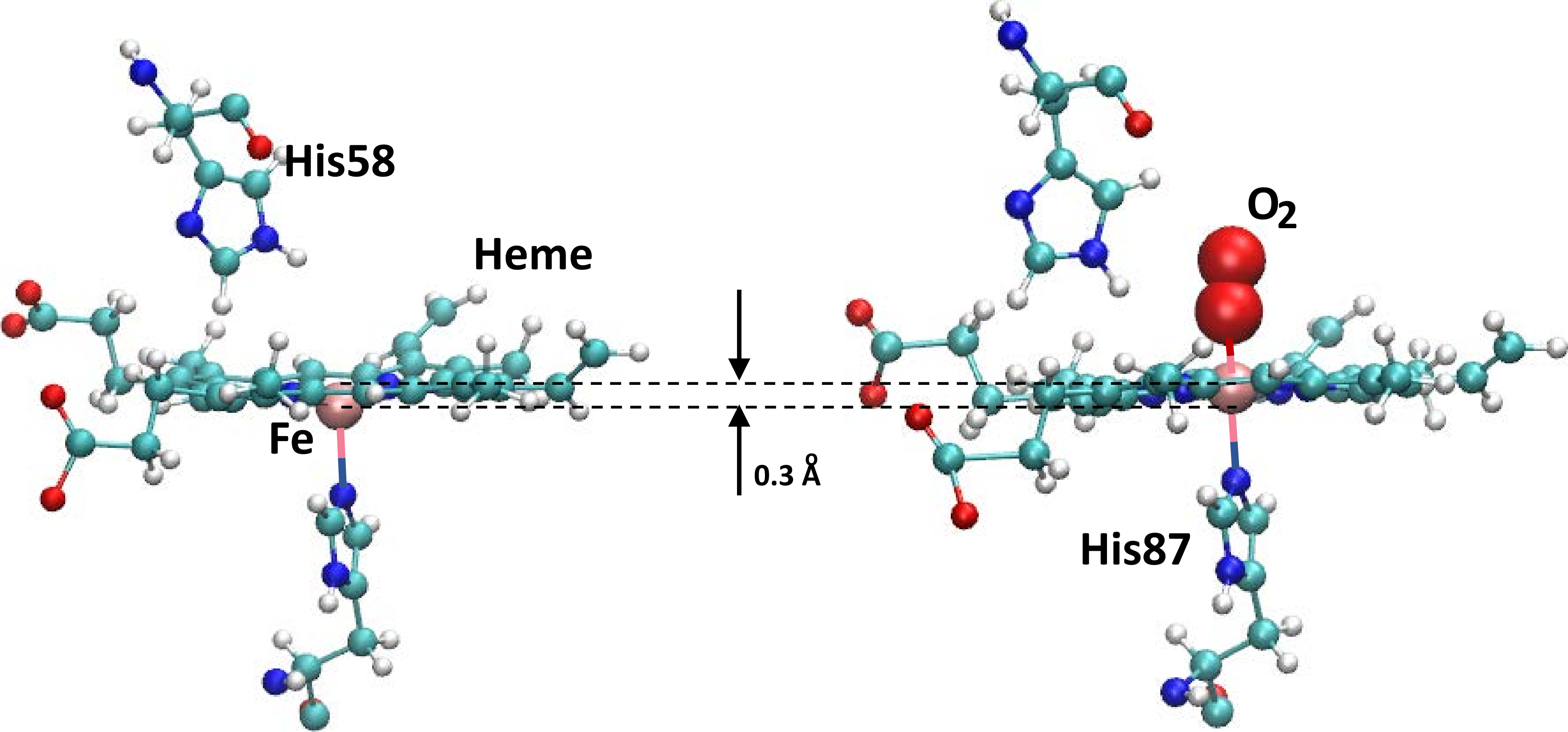}
\caption{The active site of Hemoglobin with the heme-group (CPK), the
  heme-iron (light red sphere), the O$_2$ ligand (dark red spheres),
  and the two histidine residues surrounding the active site with
  His87 covalently bound to the heme-iron. The position of the iron
  atom is $\sim 0.3$ \AA\/ below the average heme plane for the
  5-coordinate heme (left) and in the heme plane for 6-coordinate heme
  (right).}
\label{fig:heme}
\end{figure}

Hemoglobin (Hb) is one of the most widely studied proteins due to its
essential role in transporting oxygen from the lungs to the
tissues. Binding of molecular oxygen, O$_2$, at the heme-iron is the
physiologically relevant step for oxygen transport. Oxygen
homeostasis\cite{semenza:2010} at the molecular level can be described
as the dynamic equilibrium between O$_2-$bound and ligand-free Hb at
one of the four heme-iron atoms and depends on the local O$_2$
concentration. The self-regulation of oxygen homeostasis is reflected
in protein allostery,\cite{cui:2008} the capacity of proteins such as
Hb to modulate their affinity towards a physiological target (here
O$_2$) by structural adaptations upon binding or removal of another
ligand (here O$_2$) at a different binding site. In other words,
allostery is the molecular embodiment of homeostasis at the cellular
level. The two most important structural states of Hb are the deoxy
structure (T$_0$), which is stable when no ligand (subscript ``0'') is
bound to the heme-iron, and the oxy structure (R$_4$), which is stable
when each of the four heme groups have a ligand (subscript ``4''),
such as oxygen, bound to them, see Figure \ref{fig:heme} for the
structure of the active site including the heme, the surrounding
histidine residues, and the bound O$_2$ ligand (right panel). The
state with the quaternary structure of R$_4$, but with no heme-bound
ligands is the R$_0$ state. Despite strong experimental
evidence\cite{edelstein:1971} that T$_0$ is significantly more stable
than R$_0$, with an equilibrium constant of $K_{\frac{T_0}{R_0}}=6.7
\times 10^5$, molecular dynamics (MD) simulations appeared to indicate
that the R$_0$ state is more stable than T$_0$.  Specifically,
simulations started with hemoglobin in its T$_0$ state have been found
to undergo a spontaneous transition into the R$_0$ state on sub-$\mu$s
time scales.\cite{Hub_2010,yusuff:2012} Understanding the molecular
details of this discrepancy between the experimentally measured and
simulated relative stabilities of the R$_0$ and T$_0$ states is
essential for establishing the reliability of simulation-based studies
of Hb and other large biomolecules. \\

\begin{figure}[b!]
\centering
\includegraphics[width=0.6\textwidth]{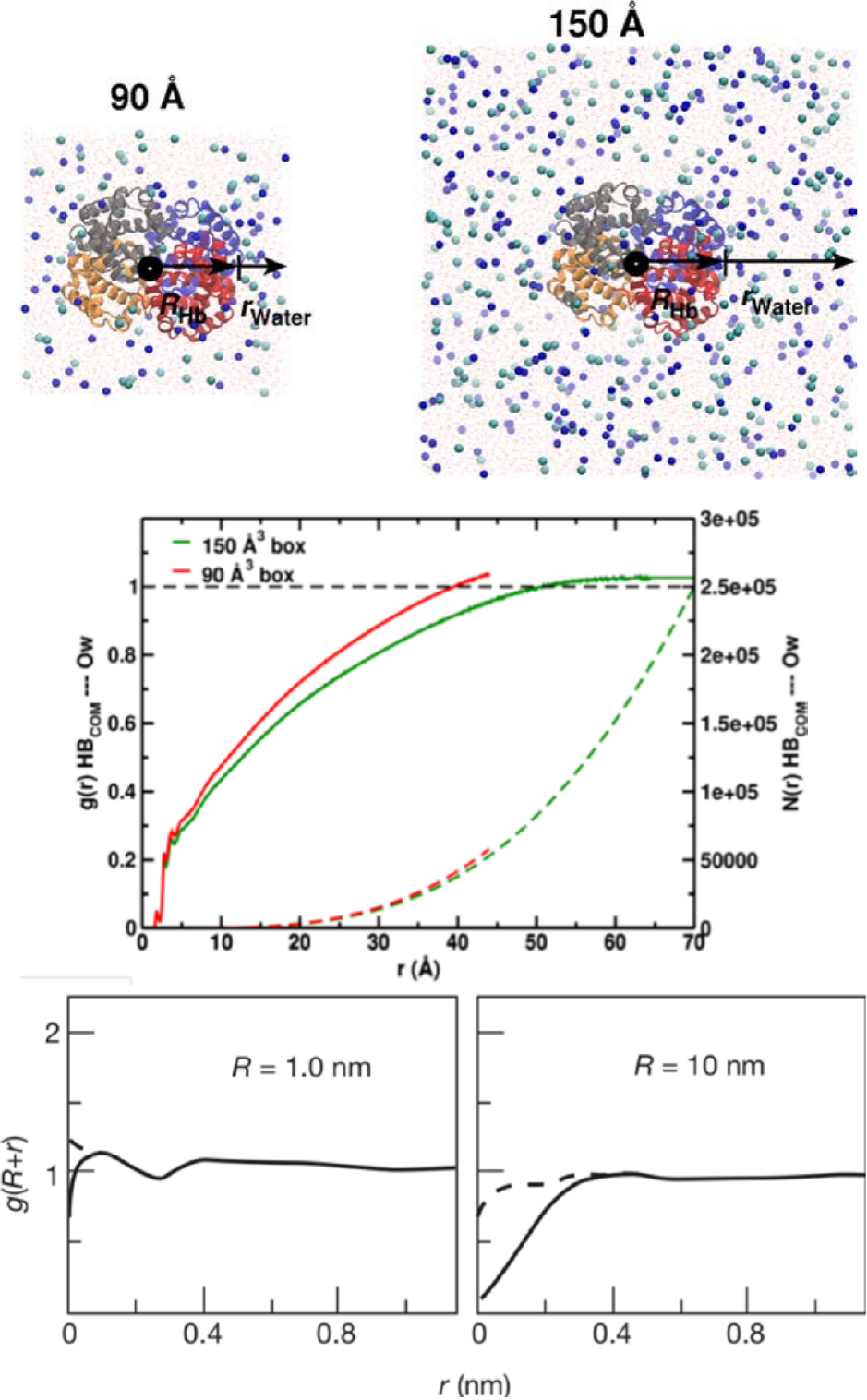}
\caption{Hydration structure depending on the size of the solute. Top:
  Hb in a 90 \AA\/ (left) and 150 \AA\/ (right) solvent box together
  with the ions (blue and cyan spheres) to give an ionic strength of
  0.15 M. Middle: solvent structure around Hb for the two different
  box sizes; the solid lines are the radial distribution function,
  $g(r)$, and the dotted lines show the total number of water
  molecules, $N(r)$, up to a distance $r$ from the surface of the
  solute. Bottom: Predictions of the solvent structure from model
  studies\cite{chandler:2005} depending on the size $R$ of the solute:
  10 \AA\/ (left) and 100 \AA\/ (right). The solid line is for an
  ideal hydrophobic solute and the dashed line includes van der Waals
  attractions between the solute and water. Figure adapted with
  permission from Ref. \cite{MM.hb:2018} with the two bottom panels
  from Ref.\cite{chandler:2005}.}
\label{fig:fig1}
\end{figure}

\noindent
In a recent set of MD simulations it was found that the T$_0$
$\rightarrow$ R$_0$ transition rate depends sensitively on the size of
the simulation box.\cite{MM.hb:2018} The simulations were initialized
with Hb in the T$_0$ state and immersed in a periodically replicated
cubic solvent box with side lengths of 75 \AA\/, 90 \AA\/, and 120
\AA\/. In such boxes transitions towards the R-state structure were
observed after 130 ns, 480 ns, and 630 ns, respectively. By contrast,
in a water box with side-length 150 \AA\/ (top row Figure
\ref{fig:fig1}), Hb remained in its T$_0$ state for the entirety of a
1.2$\mu$s simulation. Extrapolation of this trend suggests that T$_0$
is the thermodynamically stable state in this water box.\\

\noindent
The results also suggested that such a large box is required for the
hydrophobic effect, which stabilizes the T$_0$ tetramer, to be
manifested. Hydrophobicity is the tendency of a solute (here Hb) to
pack with itself and to exclude water molecules. In other words, the
solute and water segregate which leads to maximization of hydrogen
bonds between water molecules while minimizing the contact area
between the solute and water. The hydrophobic effect is one of the
main driving forces in the formation of biological interfaces,
including cell membranes and vesicles and thus is essential for
life.\cite{tanford:1978}\\

\noindent
The importance of the hydrophobic effect as an organizing ``force'' in
biological systems has been recognized for quite some
time. Hydrophobicity has been discussed and established as an
important driver in protein folding or for the compartmentalization of
cells.\cite{tanford:1978} However, the hydrophobic effect is also
prevalent in everyday life (in the function of detergents or
emulsions), in materials sciences, adhesion, and
confinement.\cite{chandler:2005} One of the findings particularly
relevant to the present discussion is the dependence of hydration of a
hydrophobe on its size, see Figure \ref{fig:fig1} bottom
row. Considering an idealized spherical cavity of radius $R$ it was
shown that the density of water a distance $r$ from the surface of the
cavity depends sensitively on its size. For a small cavity ($R \sim 4$
\AA\/, the size of methane, CH$_4$) the water density adjacent to the
cavity is larger by a factor of two than the bulk density of water
because water attempts to maintain a H-bonding network as strong and
dense as possible.\cite{chandler:2005} This changes with increasing
size of the cavity. For a cavity the size of hemoglobin ($R \sim 25$
\AA\/) solvent density around the protein is depleted and approaches
bulk density asymptotically without evident structure in the radial
distribution function, $g(r)$, see bottom Figure \ref{fig:fig1}. These
findings are consistent with earlier MD simulations of hydrophobic
hydration of melittin.\cite{rossky:1998} In these simulations it was
found that depending on the local curvature of the protein (e.g. a
``flat'' $\beta-$sheet region compared with a ``pointed'' turn),
hydrophobic residues of even the same chemical type show different
water hydration shells due to local constraints on hydrogen bonding
which lead to different orientational water structures. In other
words: flat surfaces impose different constraints on the water
H-bonding patterns than surfaces with large curvature. The predicted
overall behaviour of the radial distribution function from model
studies, as shown in the bottom panels of Figure \ref{fig:fig1}, is
consistent with those found from the atomistic simulations for Hb, see
middle panel of Figure \ref{fig:fig1}.\\

\noindent
These findings demonstrate that the global and local structure of the
solute also influence the global and local hydration which, in turn,
affect the thermodynamic stability of the solute. Because the root
mean squared difference between the T$_0$ and the R$_4$ structures of
Hb is $\sim 5$ \AA\/, their hydration is also expected to differ.
This supports the observed dependence of the thermodynamic stability
of T$_0$ vs. R$_0$, where R$_0$ is expected to be R$_4$-like, on the
size of the water box as found from the MD
simulations.\cite{MM.hb:2018} While the statistical significance of
these findings has been a topic of recent discussion in the
literature,\cite{degroot:2019,MM.hb:2019} the dynamic stability of the
T$_0$ state exhibits a clear and systematic dependence on the size of
the solvent box. The differences in the degree of hydration between
the T- and the R-states is also consistent with earlier work based on
individual X-ray structures.\cite{chothia:1985} It is well established
that the $\alpha_1 \beta_2$ and $\alpha_2 \beta_1$ interfaces in Hb
are large and closely packed. Although the allosteric transition has
been shown to be more complex,\cite{karplus.hb:2011} it can be
described as a $\sim 15^\circ$ rotation of the $\alpha_1 \beta_1$
dimer relative to the $\alpha_2 \beta_2$ dimer. The T$\rightarrow$R
transition also involves breaking of several salt bridges, in accord
with the Perutz mechanism.\cite{perutz:1970,perutz:1998}
Concomitantly, the buried surface of the R-state is reduced by $\sim
700$ \AA\/$^2$ as compared with the T-state. Based on the relationship
between solvent accessible surface area and the associated hydrophobic
contribution to the free energy,\cite{chothia:1974} burial of
additional protein surface will differentially stabilize the T-state
relative to the R-state. Hence, the decrease in the protein surface
exposed to the solvent suggests that hydrophobic effects should
stabilize the T-state. Further analysis is required to provide
conclusive evidence of the role of the hydrophobic effect and to
reveal the mechanistic origin of the dependence of the thermodynamic
stability of the T$_0$ state, relative to the R$_0$ state, on the
simulation box size. A related interesting and challenging aspect
concerns the question of whether the solvent water follows the
conformational T$\rightarrow$R transition or whether water drives this
transition. It is also possible that during different phases of this
transition the roles of solvent and solute switch or that it is more
meaningful to consider the solvent-solute system as a whole.\\

\begin{figure}[b!]
\centering
\includegraphics[width=0.75\textwidth]{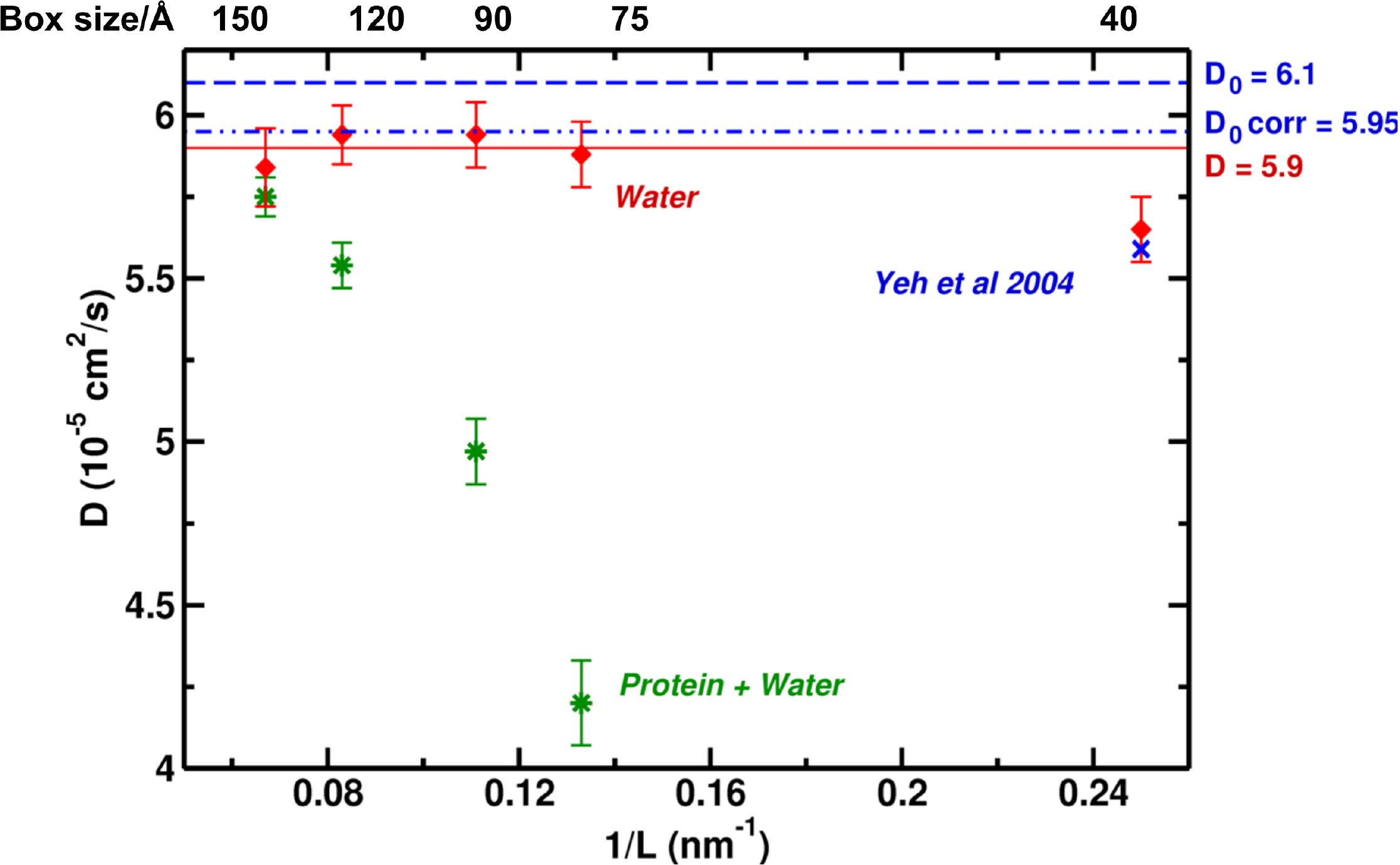}
\caption{The dependence of the water diffusivity on the size of the
  simulation system with (green symbols) and without (red symbols) Hb
  as the solute. Direct comparison for the smallest simulation box
  with that from the literature\cite{hummer:2004} is
  favourable. Figure adapted with permission from
  Ref. \cite{MM.hb:2018}.}
\label{fig:fig2}
\end{figure}

\noindent
It is of interest to note that recent work on the A$\beta$ peptide,
which is the main component of amyloid plaques involved in Alzheimer's
disease, found that different solvent box sizes yield similar radius
of gyration, secondary structure, intrapeptide, and peptide-water
hydrogen bonds.\cite{kepp:2019} However, of considerable interest for
the present discussion, it was reported that the hydrophobic surface
area and exposure of the backbone conformations depend significantly
and sensitively on the solvent box size, irrespective of the force
field used in the simulations.\cite{kepp:2019} Similarly, MD studies
of nanoparticle diffusion also reported differences in the hydration
depending on the size of the solvent box.\cite{cui:2021} For Hb as the
solute the water diffusion coefficients also depends on the size of
the simulation box, see Figure \ref{fig:fig2}. It is found that with
increasing box size the diffusivity of water for the system including
hemoglobin approaches that of the solvent alone. Finally, following a
tangential approach to the problem by use of quasichemical theory it
was reported that both hydrophilic and hydrophobic contributions to
hydration depend on system size. They are predicted to decrease with
increasing system size. The net hydration free energy benefits
somewhat from the compensation of hydrophilic and hydrophobic
contributions which is akin to entropy/enthalpy
compensation.\cite{sharp:2001,chodera:2013} Nevertheless, a large
system appears necessary to describe correctly the balance of these
contributions to the hydration of the
macromolecule.\cite{tomar:2020}\\

\noindent
More molecularly resolved and quantitative work has been done to
understand the local hydration of Hb in its T$_0$ and R$_0$
states.\cite{MM.hb:2020} For this, the local hydrophobicity (LH) was
evaluated\cite{willard:2010,willard:2018} along MD trajectories of the
two conformational states in differently sized water boxes. The local
hydrophobicity can be viewed as a generalization of a radial
distribution function in that it provides information about both the
presence {\it and} orientation of water molecules at an
interface. Such an analysis for the T$_0$ and R$_0$ states of Hb found
that the breaking and formation of salt bridges at the $\alpha_1
\beta_2$ and $\alpha_2 \beta_1$ interface is accompanied by changes in
LH.\\

\noindent
The above provides a molecular view of the Perutz mechanism which is
based on a two-state model involving an equilibrium between the T- and
the R-states.\cite{antonini:1971.chap14,perutz:1970} In the Perutz
model the T$_0$-state is ``tense'', constrained by salt bridges
between the C-termini of the four subunits and it has a low oxygen
affinity with no ligands bound, whereas the R$_4$-state is
``relaxed'', the salt bridges are broken and ligands are bound to the
heme-irons. The T$_0$-state has the (5-coordinated) heme-iron out of
plane displaced towards the proximal histidine His87, see Figure
\ref{fig:heme}, whereas in the R$_4$ state the heme-iron is in the
plane and is six-coordinated with the ligand occupying the remaining
free valence as shown in Figure \ref{fig:heme}. Every heme-iron atom
is coordinated to a histidine residue at its distal site (``below the
heme plane''). The histidine residues are His87 in the $\alpha-$chains
and His92 in the $\beta-$chains. Because these histidine residues are
part of the F-helix that is linked to the E-helix (containing the
proximal histidine residue) forming a loop, motion at the distal side
of each subunit can be efficiently transduced to other relevant parts
of the protein.\cite{bartunik:1999} The T$_0$/R$_4$ equilibrium it
thought to be governed primarily by the position of the iron atoms
relative to the porphyrin and the salt bridge stability and dynamics
is considered to be linked to the distal
histidines.\cite{perutz:1970,perutz:1998} This interplay between
ligand binding, local and global conformational changes, the change in
interface exposed to the solvent and subsequent rearrangement of
solvent leads to a logical chain of events spanning various length
scales (from atomic to mesoscopic) that govern Hb function.\\

\noindent
For Hb in cubic water boxes with 90 \AA\/ and 120 \AA\/ edge length it
was found that simulations initialized in the T$_0$ state decay to
known but different intermediate structures upon destabilization of
the $\alpha \beta$ interface following a decrease in LH; i.e. as a
consequence of reduced water density or change of water orientation at
the protein/water interface.\cite{MM.hb:2020} This is in line with
earlier simulations\cite{MM.hb:2018} that reported a reduced number of
water-water hydrogen bonds in smaller simulation boxes which shifts
the equilibrium between water-protein and water-water contacts and
changes the activity of water. Interestingly, for decreasing box
sizes, simulations for the A$\beta$ peptide reported a pronounced
increase in the hydrophobic surface area\cite{kepp:2019} and studies
of protein G found an increase in the hydrophobic contribution due to
a decrease in solvent density fluctuations as the system size
decreases.\cite{tomar:2020} This is consistent with studies for Hb
that found larger hydrophobic exposure in smaller simulation
boxes.\cite{MM.hb:2020}\\

\section{Characterization of the Protein/Water Interface from Measurements on RBCs}
In contrast to studies of individual macromolecules, such as Hb, in
aqueous solution containing adequate concentrations of anions and
cations (see previous sections), proteins in a realistic cellular
environment experience extreme crowding.\cite{minton:2008} Hence,
although the above studies are of interest to better understand the
physical behaviour of complex macromolecules at a molecular level,
they are not necessarily directly relevant to the behaviour of
proteins under physiological conditions.\\

\noindent
A notable experiment\cite{stadler2008cytoplasmic} determined the
diffusivity of water in Hb/water mixtures at protein concentrations of
$\sim 330$ mg/mL, which is representative of the crowded cellular
environment with protein concentrations of up to 400 mg/mL. Using
quasielastic incoherent neutron scattering (QENS) to probe the water
dynamics of RBCs in D$_2$O and H$_2$O buffer, the cytoplasmic dynamics
of H$_2$O was separated from membrane and macromolecular dynamics. The
difference in the two signals is dominated by the water dynamics
because hydrogen nuclei have an incoherent scattering cross section
that is $\sim 40$ times larger than that of macromolecules or
deuterium. The translational diffusion of cellular water was found to
be nearly identical to that of H$_2$O buffer which is consistent with
the results from MD simulations if sufficiently large solvent boxes
are used, see Figure \ref{fig:fig2}. However, this finding is
surprising for cellular environments for three reasons. First, the
average separation of macromolecules in a crowded cellular environment
is of the order of 10 \AA\/ which corresponds to only $\sim 3$ layers
of water molecules. Second, NMR experiments\cite{halle:2008.2} and MD
simulations\cite{laage:2012} have found that the reorientation
dynamics of water on the protein surface is slowed down by a factor of
2 to 3 compared with water in the bulk. Third, time resolved
fluorescence spectroscopy reported that a significant fraction of the
water molecules is slowed down by an order of
magnitude.\cite{zewail:2002} Although in fact, the notion that the
dynamics of water adjacent to a protein surface differs from that in
the bulk dates back at least 60 years,\cite{bernal:1965} there is as
yet no explanation for the high diffusivity of cellular water.\\

\noindent
Much effort has gone into characterizing the behaviour of Hb in
RBCs.\cite{longeville:2017} Of particular relevance is the
understanding of the sensitivity of the volume of RBCs to changes in
the osmolality of the surrounding medium. Such effects were referred
to as ``anomalous osmotic behaviour'' of RBCs.\cite{savitz:1964}
Interestingly, the molecular explanation for the apparent anomaly is a
cooperative effect by which the total charge of Hb decreases with
increased Hb concentration.\cite{gary:1968}\\

\section{Relevance of {\it In Vitro} Studies to Physiology}
As noted above, in a cellular environment the spatial separation
between proteins is of the order of 10 \AA\/. This differs
considerably from most MD studies that investigate one or a few
proteins in solution. Similarly, NMR experiments are carried out under
dilute conditions that avoid clustering of the proteins. Hence, the
question arises in what sense such {\it in vitro} studies are relevant
to the situation encountered {\it in vivo}.\\

\noindent
A first obvious difference, as already mentioned, is the ``crowding''
encountered in real cells. Typically, physico-chemical experiments aim
at {\it reducing} the complexity in order to obtain {\it specific}
information about the system of interest - here the energetics and
dynamics of Hb. However, crowding and the presence of multiple
interaction partners generally makes the interpretation of experiments
on RBCs difficult. For example, when using infrared spectroscopy to
probe site-specific dynamics in a protein much effort is spent in
finding molecules that absorb in a frequency range that is largely
devoid of responses from the protein.\cite{hamm:2015} This range
extends from $\sim 1800$ cm$^{-1}$ to $\sim 2900$ cm$^{-1}$. Hence,
spectroscopic probes such as -CN,\cite{romesberg.cn:2011}
-SCN,\cite{bredenbeck:2014} or
N$_3$\cite{hamm.aha:2012,MM.n3:2019,MM.lys:2021} which absorb between
2000 cm$^{-1}$ and 2300 cm$^{-1}$ are ideal reporters since all
signals in this frequency range can be unambiguously assigned to the
reporter groups.\\

\noindent
One significant study directly probed the water dynamics in a cellular
environment.\cite{halle:2008} Contradicting the view that a
substantial fraction of cell water is strongly perturbed, it was found
that $\sim 85$ \% of cell water in {\it E. coli} and in the extreme
halophile {\it Haloarcula marismortui} had bulk-like dynamics,
consistent with the results in Figure \ref{fig:fig2}. The remaining
$\sim 15$ \% of cell water interacts directly with biomolecular
surfaces and is motionally retarded by a factor $15 \pm 3$ on average,
corresponding to a rotational correlation time of 27 ps. This dynamic
perturbation is three times larger than for small monomeric proteins
in solution, a difference that was attributed to secluded surface
hydration sites in supramolecular assemblies.\\

\noindent
More recently, all atom MD simulations of the {\it Mycoplasma
  genitalium} (Mg), the simplest bacterium, with a genome of 470 genes
versus {\it E. coli}, which has about 4600 genes, have been
performed.\cite{feig:2016} They included all molecular components
(i.e.  proteins, RNA, metabolites, ion, and water) explicitly in
atomic detail with a total of $\sim 104$ million atoms.  The system
was simulated for 20 $\mu$s. Even with this short simulation time some
interesting results were obtained; e.g., partial denaturation due to
protein-protein interactions occurred and macromolecular diffusion was
slowed down.\\

\noindent
Another example of the differences between {\it in vivo} and {\it in
  vitro} studies was found for the production of recombinant
adeno-associated viruses.\cite{paulk:2020} Although this example is
not related to hemoglobin {\it per se}, it nicely illustrates that the
behaviour of a complex system (such as a multimeric protein) depends
on the structure and composition of its environment (i.e., in a cell
or under idealized laboratory conditions). For the production of such
viruses, all conditions were maintained identical except for the host
cell species in which they were grown. Interestingly, the post
translational modifications of the viruses expressed in the two
different cell types differed and the sites at which methylation
occurred also depended on the cell line that was used. It was
concluded that virus receptor binding, trafficking, or expression
kinetics can depend on the method used to grow the virus.\\

\section{Summary}
The present work highlights the importance of a molecular-level
understanding of the interface between hemoglobin and its environment,
notably water, under {\it in vivo} and {\it in vitro} conditions. In
cellular environments, such as RBCs, crowding is the main difference
when compared with more idealized realizations of a system {\it in
  vitro} in most computational and physico-chemical
experiments. Although following such ``divide-and-conquer'' approaches
has provided remarkable insights into the dynamics and thermodynamics
of complex systems, their relevance to those experienced under
physiological conditions needs to be examined in detail.  The finding
that 85 \% of the water in cells behave like bulk water and only 15 \%
has slowed its diffusion by one order of magnitude suggests that
laboratory experiments and molecular simulations with slight
modifications (``crowders'') should be able to emulate many aspects of
real-cell environments without unduly increasing the complexity of the
system investigated. This is even more likely to be a meaningful
approach as simulations for Hb carried out in sufficiently large
solvent boxes confirm these findings. Bridging the gap between
idealized single-molecule scenarios typically considered in current
laboratory and simulation-based experiments and the crowded cellular
environments is a challenge that will be taken up in the near future
to obtain a molecular-level understanding of complex biological
systems, including living cells.\\

\section*{Acknowledgment}
The authors thank Adam Willard for insightful comments and thoughtful
correspondence. Support by the Swiss National Science Foundation
through grants 200021-117810, the NCCR MUST (to MM), and the
University of Basel is acknowledged. The support of MK by the CHARMM
Development Project is gratefully acknowledged.\\

\bibliography{new}
\end{document}